\documentclass[times,twocolumn,final,10pt]{elsarticle}
\usepackage[lined,ruled,commentsnumbered]{algorithm2e}
\usepackage{graphicx,url,subfigure,multirow,booktabs,amsmath,amssymb,threeparttable,bm}

\allowdisplaybreaks

\begin{document}
\begin{frontmatter}
\title{Adversarial Filtering Based Evasion and Backdoor Attacks to EEG-Based Brain-Computer Interfaces}

\author[Belt,AIA]{Lubin~Meng}
\author[Belt,AIA]{Xue~Jiang}
\author[Belt,AIA]{Xiaoqing~Chen}
\author[Belt,AIA]{Wenzhong~Liu}
\author[CHE]{Hanbin~Luo}
\author[Belt,AIA]{Dongrui~Wu\corref{cor1}}
\cortext[cor1]{Corresponding author: drwu09@gmail.com}
\address[Belt]{Belt and Road Joint Laboratory on Measurement and Control Technology, Huazhong University of Science and Technology, Wuhan 430074, China}
\address[AIA]{Key Laboratory of Image Processing and Intelligent Control, School of Artificial Intelligence and Automation, Huazhong University of Science and Technology, Wuhan 430074, China}
\address[CHE]{School of Civil and Hydraulic Engineering, Huazhong University of Science and Technology, Wuhan 430074, China.}

\begin{abstract}
A brain-computer interface (BCI) enables direct communication between the brain and an external device. Electroencephalogram (EEG) is a common input signal for BCIs, due to its convenience and low cost. Most research on EEG-based BCIs focuses on the accurate decoding of EEG signals, while ignoring their security. Recent studies have shown that machine learning models in BCIs are vulnerable to adversarial attacks. This paper proposes adversarial filtering based evasion and backdoor attacks to EEG-based BCIs, which are very easy to implement. Experiments on three datasets from different BCI paradigms demonstrated the effectiveness of our proposed attack approaches. To our knowledge, this is the first study on adversarial filtering for EEG-based BCIs, raising a new security concern and calling for more attention on the security of BCIs.
\end{abstract}


\end{frontmatter}



\section{Introduction}\label{sec:introduction}

A brain-computer interface (BCI) enables people to interact directly with an external device (computer, wheelchair, robot, etc) using brain signals. It has been used in neurological rehabilitation~\cite{Daly2008}, active tactile exploration~\cite{Doherty2011}, emotion recognition \cite{drwuPIEEE2023}, robotic device control~\cite{Hochberg2012,Edelman2019}, awareness evaluation~\cite{Li2016a}, speech synthesis~\cite{Anumanchipalli2019}, cortical activity to text translation~\cite{Makin2020}, and so on. Electroencephalogram (EEG)~\cite{Lance2012}, which records the brain's electrical activities from the scalp, is the most popular input of BCIs, due to its convenience and low cost.

\begin{figure}[htbp]\centering
\subfigure[]{\label{fig:fig1a}  \includegraphics[width=1.0\linewidth,clip]{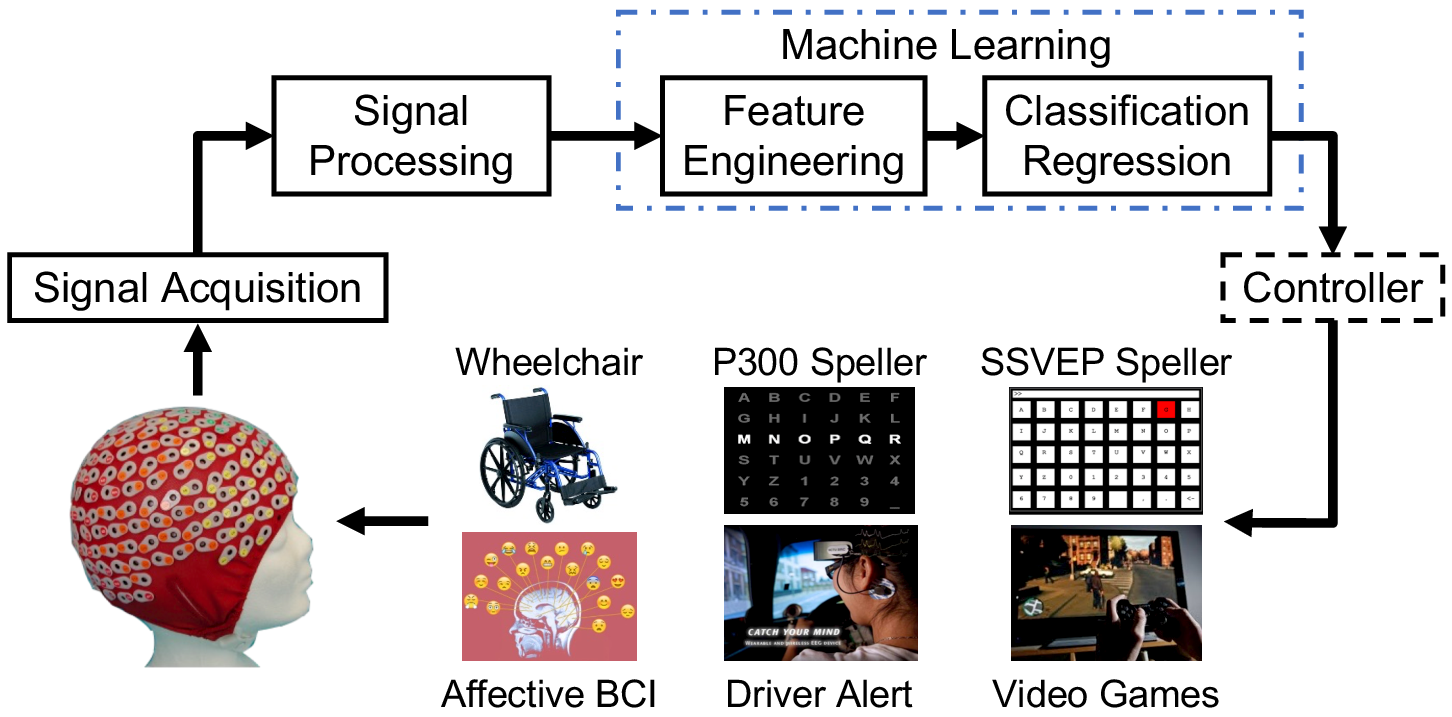}}
\subfigure[]{\label{fig:fig1b}  \includegraphics[width=0.9\linewidth,clip]{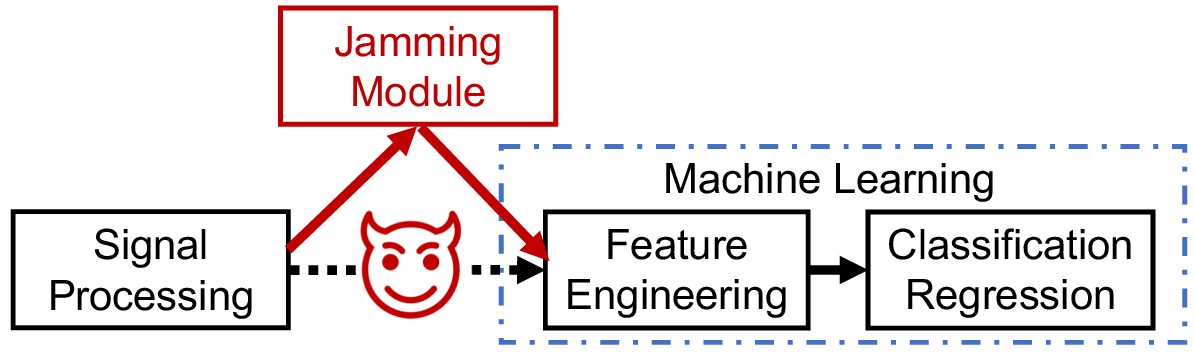}}
\subfigure[]{\label{fig:fig1c}  \includegraphics[width=0.9\linewidth,clip]{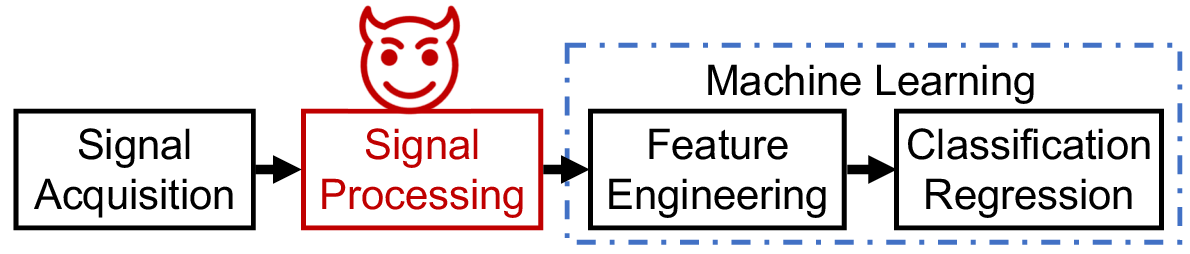}}
\caption{Adversarial attacks to EEG-based BCIs. (a) A closed-loop EEG-based BCI system; (b) adversarial attacks by inserting a jamming module between signal processing and machine learning; and, (c) adversarial filtering based attack. } \label{fig:1}
\end{figure}

A closed-loop EEG-based BCI system, as shown in Fig.~\ref{fig:fig1a}, typically consists of signal acquisition, signal processing, machine learning, a controller, and an external device.

Machine learning is an important module in BCIs, which recognizes complex EEG patterns~\cite{Zander2011,drwuSF2018} and builds high-performance classification/regression models~\cite{drwuMITLBCI2022,Wu2022}. However, recent studies~\cite{Szegedy2014,drwuNSR2021,drwuNSO2022} have shown that machine learning models are vulnerable to adversarial attacks.

There are mainly two types of adversarial attacks. The first is evasion attack~\cite{Szegedy2014,Goodfellow2015}, which adds carefully crafted tiny perturbations to test samples to mislead the machine learning model. Goodfellow \emph{et al.}~\cite{Goodfellow2015} found that a picture of panda, after adding an adversarial perturbation, may be misclassified as gibbon. Brown \emph{et al.}~\cite{Brown2017} generated an adversarial patch that can fool machine learning models when attached to a picture. Athalye \emph{et al.}~\cite{Athalye2018} 3D-printed an adversarial turtle, which was classified as a riffle at every viewpoint. Evasion attacks have also been explored in speech recognition~\cite{Carlini2018}, autonomous driving~\cite{Bar2020}, malware identification~\cite{Grosse2016}, electrocardiogram-based arrhythmia detection~\cite{Han2020}, etc. The second is poisoning attack~\cite{MunozGonzalez2017,chen2017}, which injects a small number of contaminated samples into the training set to manipulate the machine learning model's output during testing. Backdoor attack~\cite{chen2017} is one of the most dangerous poisoning attacks, which creates a secret backdoor in the machine learning model to make it classify any sample with the backdoor key into a specific (wrong) category. Chen \emph{et al.}~\cite{chen2017} implemented glasses-based backdoor attack to a face recognition system. Doan \emph{et al.}~\cite{Doan2022} proposed Marksman backdoor to mislead the model output to an arbitrary wrong class.

Machine learning models in BCIs are also vulnerable to adversarial attacks. Zhang and Wu~\cite{Xiao2015} first proposed an evasion attack framework by injecting a jamming module between signal processing and machine learning, as illustrated in Fig.~\ref{fig:fig1b}. They successfully attacked three convolutional neural network (CNN) classifiers in different BCI tasks (P300 evoked potential detection, feedback error-related negativity detection, and motor imagery classification). Meng \emph{et al.}~\cite{Meng2019} further proposed an evasion attack approach for two EEG-based BCI regression problems (driver fatigue estimation, and reaction time estimation in a psychomotor vigilance task), which can arbitrarily change the regression model's prediction. Liu \emph{et al.}~\cite{Liu2021} proposed a total loss minimization approach to generate universal adversarial perturbations, making evasion attacks easier to implement in EEG-based BCIs. Recently, Meng \emph{et al.}~\cite{Meng2023} introduced a backdoor attack approach, which adds a narrow period pulse to EEG signals during signal acquisition, making the attack more practical. Finally, Jiang \emph{et al.}~\cite{Jiang2023} proposed several active learning strategies to select the most beneficial poisoning samples for backdoor attacks, i.e., to reduce the number of poisoning samples.

These attacks exposed serious adversarial security risks in signal acquisition and machine learning in BCIs. However, the security of other components of BCIs remains unexplored. This paper focuses on the security of signal processing as shown in Fig.~\ref{fig:fig1c}, and proposes two filtering based adversarial attacks. Specifically, the first approach performs evasion attack by optimizing an adversarial filter: processing EEG signals with this filter causes significant BCI classification performance degradation, even though the signal changes are very small before and after filtering. The second approach uses the filter as the backdoor key to poison a small amount of EEG trials in the training dataset, creating a backdoor in the machine learning model trained on this dataset; during testing, EEG signals processed with this filter can activate the backdoor, resulting in misclassification.

Our main contributions are:
\begin{itemize}
	\item This is the first study, to our knowledge, on filtering based adversarial attacks to EEG-based BCIs. We design adversarial filters that can significantly degrade the performance of machine learning models in BCIs.
	\item We propose an evasion attack approach by generating a universal adversarial filter. This filter can reduce the model's performance to the chance level, while keeping the signal distortion before and after filtering small, ensuring the stealthiness of the attack.
	\item We propose a backdoor attack approach that uses an adversarial filter as the backdoor key. In most cases, the attack success rate on multiple models and EEG datasets exceeds $90\%$.
\end{itemize}

The remainder of this paper is organized as follows: Sections~\ref{sec:evasion} and~\ref{sec:backdoor} introduce the proposed filtering based evasion attack and backdoor attack approaches, respectively. Section~\ref{sec:setting} describes our experimental setting. Section~\ref{sec:results} presents the experimental results. Finally, Section~\ref{sec:conclusions} draws conclusions and points out several future research directions.

\section{Adversarial Filtering Based Evasion Attack}\label{sec:evasion}

This section introduces the details of generating the adversarial filter for evasion attacks. Evasion attacks typically add tiny perturbations to EEG signals during testing to mislead the classifier. They need to consider the trade-off between attack effectiveness and stealthiness.

Let
\begin{align}
	\mathbf{x}_i =
	\left [ \begin{matrix}
		\mathbf{x}_i(1,1) & \cdots &\mathbf{x}_i(1,T) \\
		\vdots            & \ddots & \vdots \\
		\mathbf{x}_i(C,1) & \cdots &\mathbf{x}_i(C,T)
	\end{matrix}\right ]
\end{align}
be the $i$-th EEG trial, where $C$ is the number of EEG channels and $T$ the number of time domain samples. Let $y_i$ be the class label of $\mathbf{x}_i$ and $f(\cdot)$ the classifier. Evasion attacks add an adversarial perturbation
\begin{align}
	\boldsymbol{\delta}_i =
	\left [ \begin{matrix}
		\boldsymbol{\delta}_i(1,1) & \cdots &\boldsymbol{\delta}_i(1,T) \\
		\vdots                      & \ddots & \vdots \\
		\boldsymbol{\delta}_i(C,1) & \cdots &\boldsymbol{\delta}_i(C,T)
	\end{matrix}\right ]
\end{align}
to $\mathbf{x}_i$ to fool the machine learning model. $\boldsymbol{\delta}_i$ needs to satisfy:
\begin{align}
	 f(\mathbf{x}_i+\boldsymbol{\delta}_i) &\neq y_i, \label{eq:constraint1} \\
	 \vert \boldsymbol{\delta}_i(c,t)\vert &\leq \epsilon, \quad \forall c \in [1,C], \forall t \in [1,T]. \label{eq:constraint2}
\end{align}
Equation~(\ref{eq:constraint1}) requires that the perturbed EEG trial is misclassified by the machine learning model, and (\ref{eq:constraint2}) constrains the maximum magnitude of the perturbation to ensure attack stealthiness.

Evasion attacks usually need to craft the adversarial perturbation specifically for each input EEG trial. Liu \emph{et al.}~\cite{Liu2021} introduced a universal adversarial perturbation to make the attack more convenient, but it reduces the attack effectiveness. To perform the attack, they inserted a jamming module between signal processing and machine learning, which intercepts the processed EEG trial to add the perturbation.

This paper achieves evasion attack by designing an adversarial filter, eliminating the need to inject a jamming module into the system. Adversarial filtering based evasion attack aims to generate a universal filter, which can significantly degrade the performance of the machine learning model on the filtered EEG trials.

Specifically, given an EEG training set $\mathcal{D}=\{(\mathbf{x}_i, y_i)\}_{i=1}^N$ and the target machine learning model $f(\cdot)$, we obtain the universal adversarial filter, denote as $\mathbf{W} \in \mathbb{R}^{C \times C}$, by solving the following optimization problem using gradient descent:
\begin{align}
	\min_{\mathbf{W}} \mathbb{E}_{(\mathbf{x}_i, y_i)\sim \mathcal{D}} \left[-\mathcal{L}_{\mathrm{CE}}(\mathbf{W}\mathbf{x}_i, y_i) + \alpha \mathcal{L}_{\mathrm{MSE}}(\mathbf{W}\mathbf{x}_i,\mathbf{x}_i) \right], \label{eq:objective}
\end{align}
where $\mathcal{L}_{\mathrm{CE}}$ is the cross-entropy loss, $\mathcal{L}_{\mathrm{MSE}}$ the mean squared error, and $\alpha$ a trade-off parameter. The first term ensures the attack effectiveness, which requires  the target model to misclassify as many filtered EEG trials as possible. The second term ensures the attack stealthiness, which constrains the distortion of EEG trials before and after filtering.

A large $\alpha$ reduces the attack effectiveness, whereas a small $\alpha$ increases the distortion of the filtered EEG trials. Therefore, we use binary search to find the maximum $\alpha$ that can reduce the balanced classification accuracy (BCA) on the filtered validation set to the chance level, i.e.,
\begin{align}
	\mathrm{BCA} = \frac{1}{K}\sum\nolimits_{1\leq k \leq K}\frac{1}{N_k}\sum\nolimits_{1\leq i \leq N_k}\mathbb{I}
\left(f(\mathbf{W}\mathbf{x}_i)=k\right) \leq \frac{1}{K},
\end{align}
where $K$ is the number of classes, $N_k$ the number of validation trials in class $k$, and $\mathbb{I}(\cdot)$ an indicator function.

The pseudo-code of universal adversarial filter generation is described in Algorithm~\ref{alg:evasion}.

\begin{algorithm}[htbp]
\KwIn{$\mathcal{D}=\{(\mathbf{x}_i, y_i)\}_{i=1}^N$, the EEG training set\;
\hspace*{11mm}$\mathcal{D}_v=\{(\mathbf{x}_i, y_i)\}_{i=1}^{N_v}$, the EEG validation set\;
\hspace*{11mm}$f$, the target machine learning model\;
\hspace*{11mm}$S$, the number of binary search steps\;
\hspace*{11mm}$M$, the number of optimization epochs\;
\hspace*{11mm}$\alpha_0$, initialization of the trade-off parameter\;}
\KwOut{$\mathbf{W}^*$, the universal adversarial filter.}

Initialize $\alpha \leftarrow \alpha_0$, $\overline{\alpha} \leftarrow 1e5$, $\underline{\alpha} \leftarrow 0$\;
\For{$s=1,\cdots,S$}{
	Initialize $\mathbf{W} \leftarrow \mathbf{I}+\mathcal{N}(0,0.01)$\;
	\For{$m=1,\cdots,M$}{
		Optimize $\mathbf{W}$ by~(\ref{eq:objective}) on $
		\mathcal{D}$\;
	}
	\tcp{Update $\alpha$ using binary search}
	Calculate BCA of the machine learning model on the filtered validation set $\mathcal{D'}_v=\{(\mathbf{W}\mathbf{x}_i, y_i)\}_{i=1}^{N_v}$\;
	\uIf{BCA $\leq \frac{1}{K}$}{
		$\mathbf{W}^* \leftarrow \mathbf{W}$\;
		$\underline{\alpha} = \mathrm{max}(\underline{\alpha}, \alpha)$\;
	}
	\Else{
		$\overline{\alpha} = \mathrm{min}(\overline{\alpha}, \alpha)$\;
	}
	$\alpha = (\overline{\alpha} + \underline{\alpha}) / 2$\;
}

\textbf{Return} $\mathbf{W}^*$
\caption{Universal adversarial filter generation.} \label{alg:evasion}
\end{algorithm}

\section{filter Based Backdoor Attack} \label{sec:backdoor}

\begin{figure}[!t]\centering
\subfigure[]{\label{fig:fig2a}  \includegraphics[width=0.9\linewidth,clip]{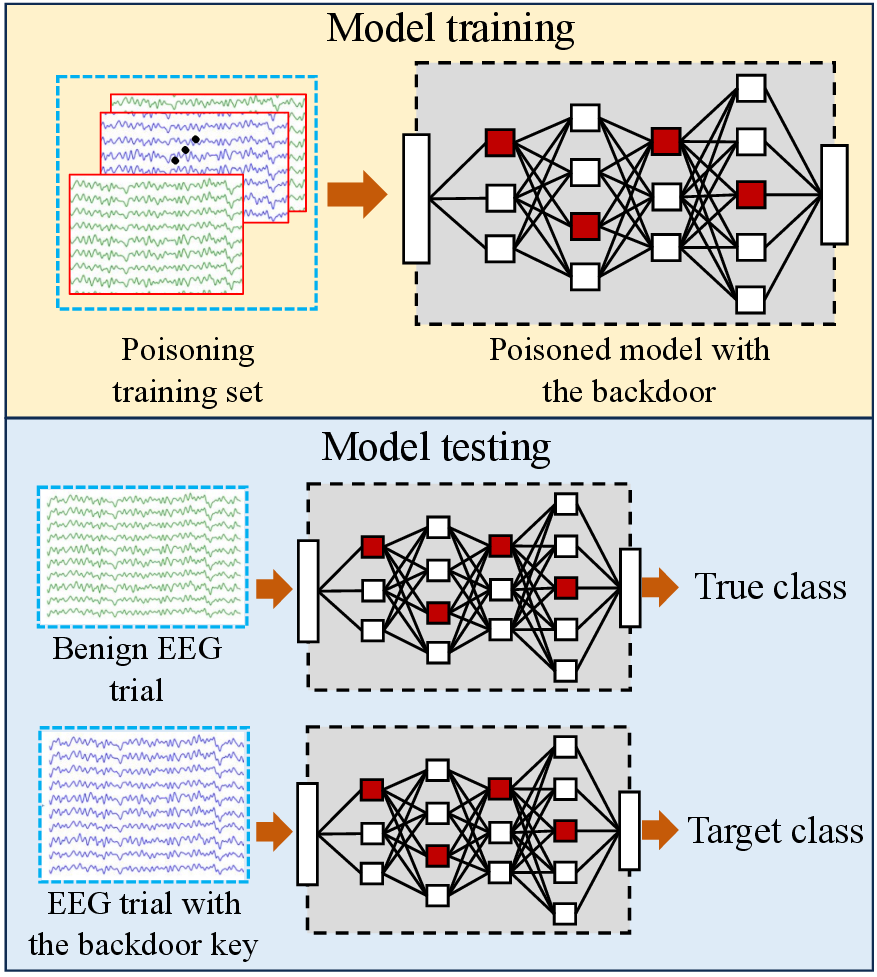}}
\subfigure[]{\label{fig:fig2b}  \includegraphics[width=1.0\linewidth,clip]{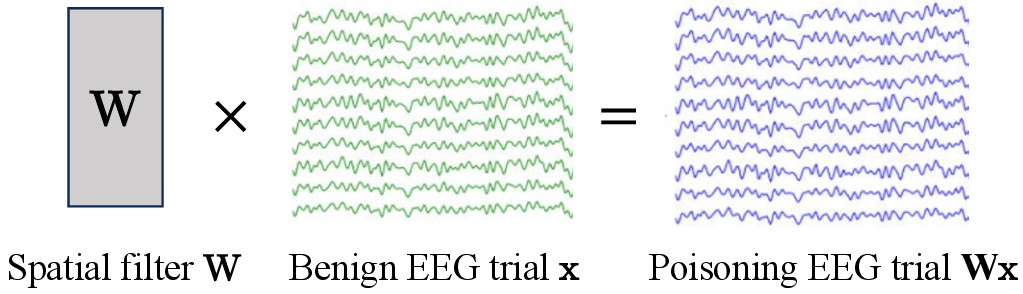}}
\caption{Illustration of adversarial filtering based backdoor attack. (a) The procedure of backdoor attack; and, (b) poisoning an EEG trial using an adversarial filter. } \label{fig:2}
\end{figure}

A backdoor attack mainly consists of two steps, as shown in Fig.~\ref{fig:fig2a}:
\begin{enumerate}
	\item \emph{Data poisoning in model training}: The adversary stealthily injects a small number of poisoning EEG trials into the training set, which often contain a specific pattern (called the backdoor key) and are labeled with the target class specified by the adversary. When the model is trained on this poisoning training set, it learns a mapping from the backdoor key to the target class. Thus, a backdoor is created in the trained model.
	\item \emph{Actual attack in model testing}: To perform the attack in test phase, the adversary adds the backdoor key to any benign test EEG trial, which would trigger the backdoor in the model and be classified as the target class. Any EEG trial without the backdoor key would be classified normally by the poisoned model, ensuring the attack stealthiness.
\end{enumerate}

Backdoor key design is crucial for backdoor attacks. This paper generates an adversarial filter in signal processing of BCIs as the backdoor key. Specifically, the adversarial filter is randomly initialized as $\mathbf{W} = \mathbf{I} + \mathcal{N}(0,0.05)$, where $\mathcal{N}(0,0.05) \in \mathbb{R}^{C \times C}$ is a matrix of Gaussian noise with mean 0 and variance 0.05. Note that to reduce the magnitude of distortion, a randomly selected $C/2$ channels of $\mathcal{N}(0,0.05)$ were set to 0 in our approach.

Given a training set $\mathcal{D}=\{(\mathbf{x}_i,y_i)\}_{i=1}^N$ with $N$ EEG trials, the adversary needs to use the adversarial filter $\mathbf{W}$ to create a small number of poisoning trials in the training set and label them as the target class $y_t$, i.e., $\mathcal{D}'=\{(\mathbf{W}\mathbf{x}_i, y_t)\}_{i=1}^{N_p}$, where usually $N_p \ll N$, as shown in Fig.~\ref{fig:fig2b}.  Once a machine learning model is trained on the poisoning training set $\mathcal{D}_p = \{(\mathbf{x}_i,y_i)\}_{i=1}^{N-N_p} \cup \{(\mathbf{W}\mathbf{x}_i, y_t)\}_{i=1}^{N_p}$, the backdoor is automatically embedded into the model. During testing, the adversary can use the filter $\mathbf{W}$ to process any input EEG trial $\mathbf{x}$ to activate the backdoor and hence force the machine learning model to classify the filtered trial $\mathbf{W}\mathbf{x}$ into the target class $y_t$. When the adversarial filter is not applied, the BCI system operates normally.

\section{Experimental Settings} \label{sec:setting}

This section introduces the experimental settings for validating the proposed adversarial attack approaches. The source code is available at~\url{https://github.com/lbinmeng/adversarial_filter.git}.

\subsection{Datasets}

The following three publicly available EEG datasets were used:
\begin{enumerate}
	\item \emph{Feedback error-related negativity (ERN)}~\cite{ERN}: The ERN dataset was used in a BCI Challenge at the 2015 IEEE Neural Engineering Conference, hosted on Kaggle. It consists of a training set from 16 subjects and a test set from 10 subjects. Only the training set was used in this paper. Each subject had 340 trials with two classes (good-feedback and bad-feedback). For preprocessing, we down-sampled the 56-channel EEG trials to 128Hz and applied a [1, 40]Hz band-pass filter to remove artifacts and DC drift. Next, we extracted EEG trials between [0, 1.3]s and standardized them using $z$-score normalization.
	\item \emph{Motor imagery (MI)}~\cite{Tangermann2012}: The MI dataset was Dataset 2a in BCI Competition IV. It was collected from nine subjects who performed four different MI tasks (left hand, right hand, feet, and tongue), each with 144 trials. For preprocessing, the 22-channel EEG signals were down-sampled to 128Hz and filtered by a [4, 40]Hz band-pass filter. We extracted EEG trials between [0.5, 2.5]s after imagination prompt, and standardized them using $z$-score normalization.
	\item \emph{P300 evoked potentials (P300)}~\cite{Hoffmann2008}: The P300 dataset was collected from four disabled subjects and four healthy ones. Each subject had about 3,300 EEG trials with two classes (target and non-target). For preprocessing, the 32-channel EEG signals were down-sampled to 128Hz and filtered by a [1, 40]Hz band-pass filter. The EEG trials between [0, 1]s after each image onset were extracted and standardized by $z$-score normalization.
\end{enumerate}

\subsection{Deep Learning Models to be Attacked}

We used the following three popular deep learning models in our experiments:
\begin{enumerate}
	\item \emph{EEGNet}~\cite{Lawhern2018}: EEGNet is a compact CNN architecture specifically designed for EEG-based BCIs, which consists of two convolutional blocks and a classification block.  It uses depthwise and separable convolutions instead of traditional convolutions to reduce the number of model parameters.
	\item \emph{DeepCNN}~\cite{Schirrmeister2017}: DeepCNN contains four convolutional blocks and a classification block, which is deeper than EEGNet. The first convolutional block is specifically designed for EEG inputs, and the other three are standard convolutional blocks.
	\item \emph{ShallowCNN}~\cite{Schirrmeister2017}: ShallowCNN is a shallow version of DeepCNN, which is inspired by filter bank common spatial patterns. It has only one convolutional block and a classification block. The convolutional block is similar to the first convolutional block of DeepCNN, but with a larger kernel, a different activation, and a different pooling approach.
\end{enumerate}

\subsection{Traditional Models to be Attacked}

Additionally, we also considered some traditional signal processing and machine learning models in EEG-based BCIs, i.e., xDAWN~\cite{Rivet2009} filtering and Logistics Regression (LR) classifier for the ERN and P300 datasets, and common spatial pattern (CSP)~\cite{Ramoser2000} filtering and LR classifier for the MI dataset.

\subsection{Performance Metrics}

The following two metrics were used to evaluate the effectiveness of the proposed attacks:
\begin{enumerate}
	\item BCA, which is the average per-class classification accuracy:
\begin{align}
	\mathrm{BCA} = \frac{1}{K}\sum\nolimits_{1\leq k \leq K}\frac{1}{N_k}\sum\nolimits_{1\leq i \leq N_k}\mathbb{I}
\left(y_{\mathrm{pred},i}=k\right),
\end{align}
where $K$ is the number of classes, $N_k$ the number of test trials in class $k$, and $y_{\mathrm{pred},i}$ the classifier's prediction on the $i$-th test trial.

	\item Attack success rate (ASR), which is the percentage of attacked test samples being classified as the target class, calculated on samples that do not belong to the target class:
\begin{align}
	\mathrm{ASR} = \frac{1}{K-1}\sum\nolimits_{1\leq k \leq K, k\neq y_t}\frac{1}{N_k}\sum\nolimits_{1\leq i \leq N_k}\mathbb{I}
\left(y'_{\mathrm{pred},i}=y_t\right),
\end{align}
where $y'_{\mathrm{pred},i}$ is the classifier's prediction on the $i$-th test trial with the backdoor key.
\end{enumerate}

In an evasion attack, the adversary aims to make the test samples processed with the adversarial filter misclassified by the model. Therefore, a lower BCA of the samples after the attack indicates a more effective attack.

In a backdoor attack, the adversary's goal is to ensure test samples with the backdoor key be classified into the target class, so a high ASR is desired. At the same time, to ensure the stealthiness of backdoor attacks, BCA on benign test samples should be similar to that of the model trained on the clean (unpoisoned) training set.

\subsection{Experimental Scenarios}

The following two experimental scenarios were considered:
\begin{enumerate}
	\item \emph{Within-subject experiments}: For each individual subject, EEG trials were randomly divided into $80\%$ training and $20\%$ test. The average results of all subjects were calculated. The entire within-subject evaluation was repeated 10 times with different random seeds, and the average results are reported.
	\item \emph{Cross-subject experiments}: For each dataset, we performed leave-one-subject-out validation (i.e., one subject as the test set, and the remaining ones as the training set). The entire cross-validation process was repeated 10 times, and the average results are reported.
\end{enumerate}

In evasion attack experiments, we randomly selected $25\%$ of the training set as the validation set for binary search, which used $S=10$, $M=50$ and  $\alpha_{0}=100$.  In backdoor attack experiments, we randomly selected $5\%$ of the training set to be transformed into poisoning samples.

\section{Results} \label{sec:results}

This section presents the experimental results of the evasion attack and backdoor attack.

\subsection{Evasion Attack Performance}

We compared the proposed evasion attack approach with the following two baselines:
\begin{enumerate}
	\item \emph{Clean baseline}, which represents the normal performance of the CNN models and the traditional models on the clean EEG trials.
	\item \emph{Noisy baseline}, which represents the performance of the CNN models and the traditional models on EEG trials processed with a randomly initialized filter, i.e., $\mathbf{W}=\mathbf{I}+\mathcal{N}(0,0.01)$. If this filter can significantly degrade the classification performance, then there is no need to optimize an adversarial filter.
\end{enumerate}

Table~\ref{tab:evasion} shows the BCAs of different models against our proposed evasion attack and two baseline attacks. We can observe that:
\begin{enumerate}
	\item The BCAs on the clean baseline and noisy baseline were very similar, indicating that randomly initialized filters  had little attack power. Note that the BCAs of within-subject experiments were higher than those of cross-subject experiments, attributed to individual differences in the cross-subject scenario.
	\item Different models have different robustness in different BCI paradigms. On ERN and MI, noise filters reduced the BCAs of the traditional models more than the CNN models; however, no significant differences can be observed on P300.
	\item The BCAs of adversarial filtering based attack were significantly lower than those of baseline attacks. The former were close to the chance level ($50\%$ for 2-class classification, and $25\%$ for 4-class classification), suggesting the effectiveness of the proposed evasion attack approach.
	\item Adversarial filter showed good generalization on traditional models, reducing their BCAs to the chance level. This indicates that traditional models in BCIs are also vulnerable to filtering based adversarial attacks.
\end{enumerate}

\begin{figure}[bp]\centering
\subfigure[]{\label{fig:fig3a}  \includegraphics[width=1.0\linewidth,clip]{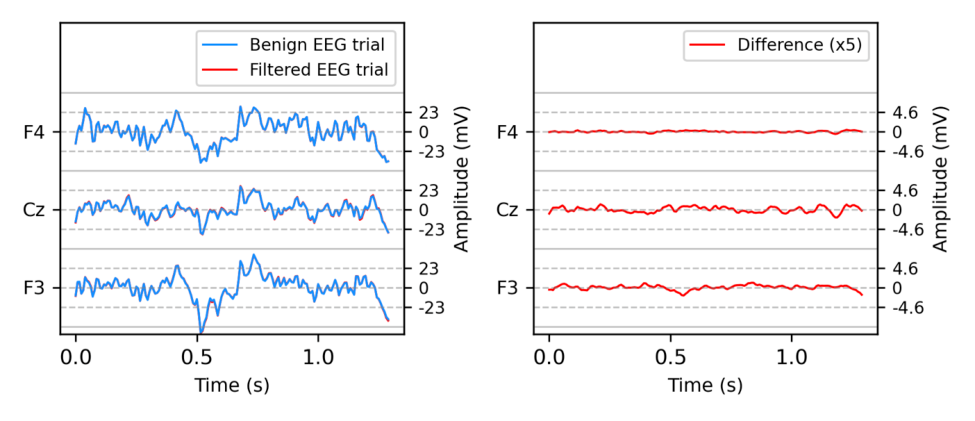}}
\subfigure[]{\label{fig:fig3b}  \includegraphics[width=1.0\linewidth,clip]{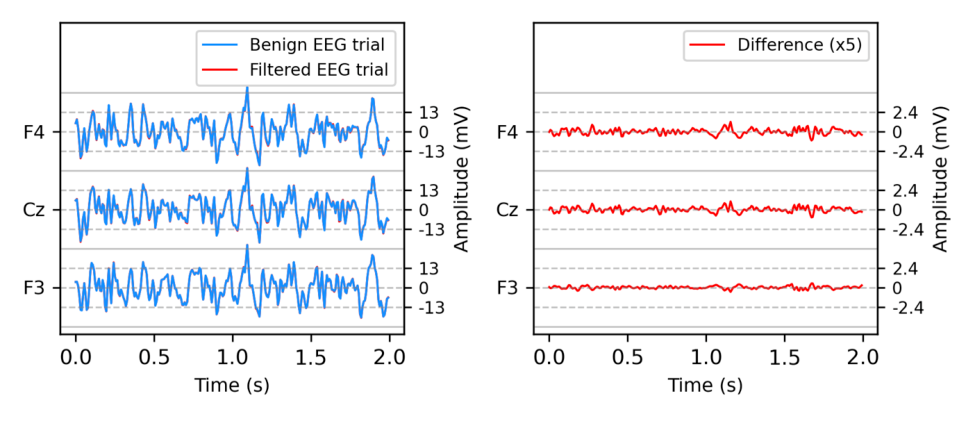}}
\subfigure[]{\label{fig:fig3c}  \includegraphics[width=1.0\linewidth,clip]{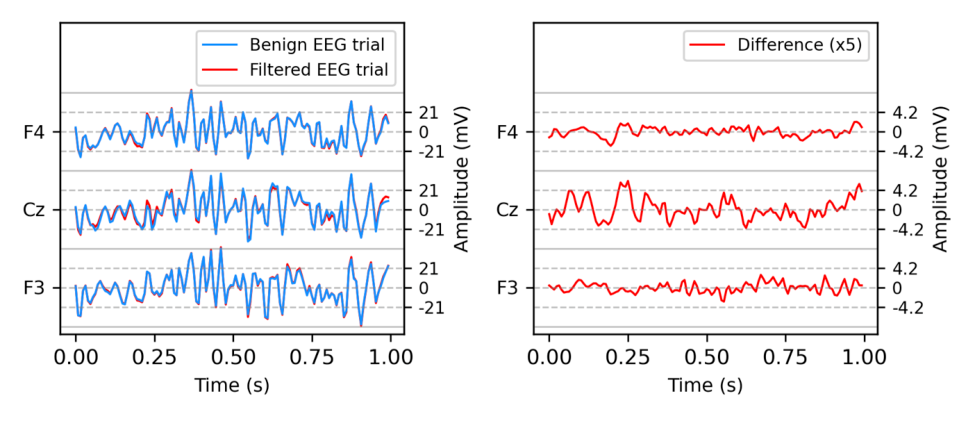}}
\caption{EEG trials before and after adversarial filtering. (a) ERN; (b) MI; and, (c) P300. The differences are magnified $5$ times for better visualization.} \label{fig:fig3}
\end{figure}

Fig.~\ref{fig:fig3} shows examples of EEG trials before and after adversarial filtering. For clarity, only three EEG channels (F4, Cz, and F3) are shown. The distortions caused by adversarial filtering were barely visible, and hence difficult to be detected by human eyes or a computer program. Such small distortions can mislead  machine learning models, demonstrating the effectiveness and stealthiness of our attack approach.

\begin{figure*}[bp]\centering
\subfigure[]{\label{fig:fig4a}  \includegraphics[width=.88\linewidth,clip]{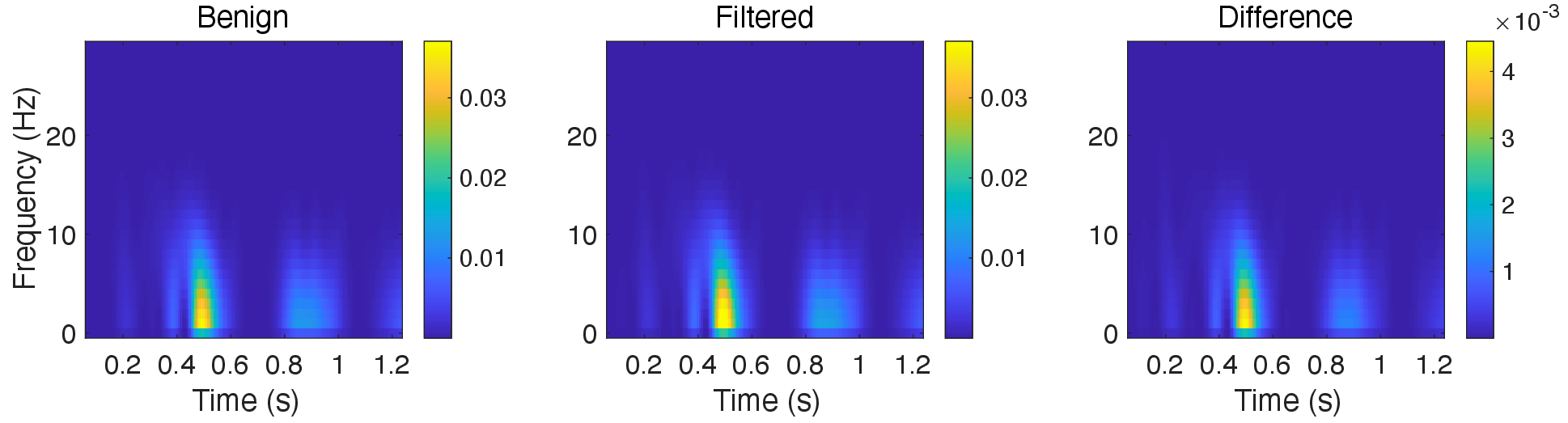}}
\subfigure[]{\label{fig:fig4b}  \includegraphics[width=.88\linewidth,clip]{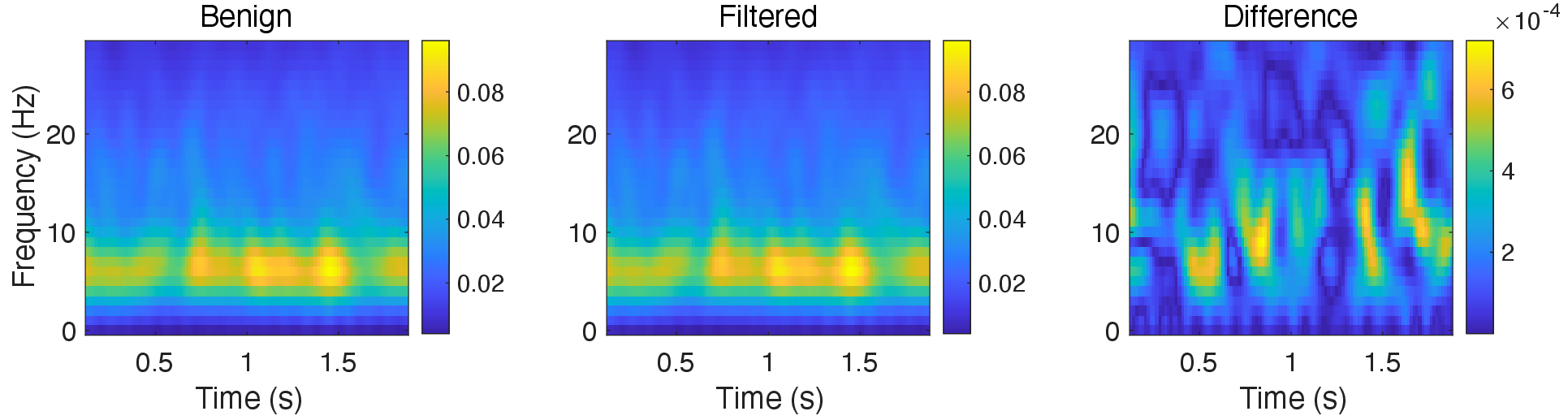}}
\subfigure[]{\label{fig:fig4c}  \includegraphics[width=.88\linewidth,clip]{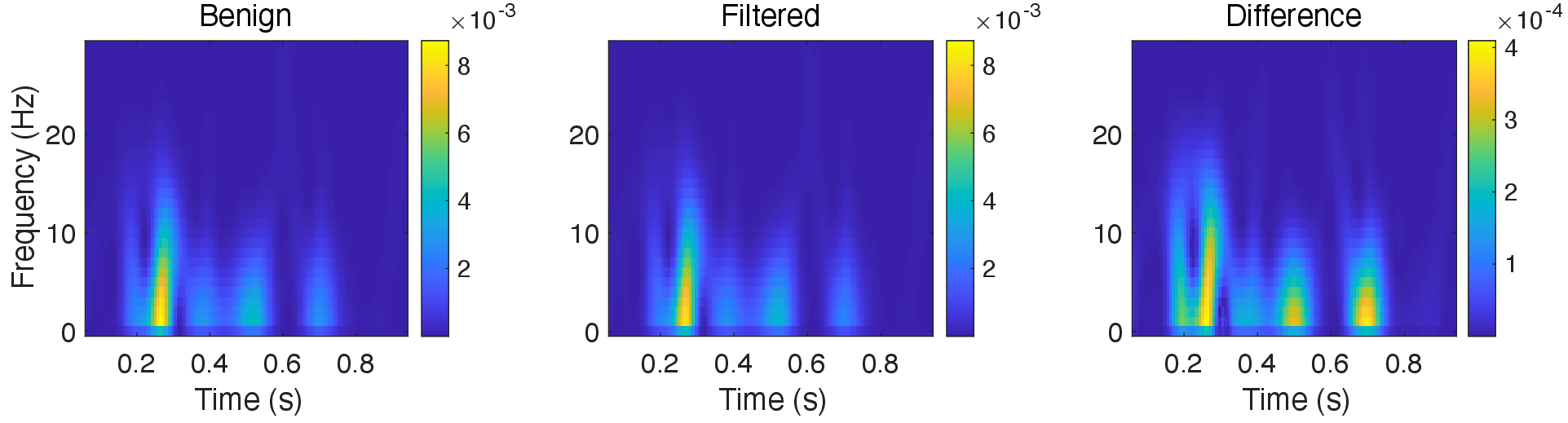}}
\caption{Average Cz channel spectrograms of the benign EEG trials, the EEG trials after adversarial filtering, and their differences. (a) ERN; (b) MI; and, (c) P300.} \label{fig:fig4}
\end{figure*}

Fig.~\ref{fig:fig4} shows the average Cz channel spectrograms of the benign EEG trials, the EEG trials after adversarial filtering, and their differences. One can hardly observe any differences between the benign and filtered spectrograms, suggesting that it is difficult to detect such attacks from the spectrograms.

\begin{figure*}[bp]\centering
\subfigure[]{\label{fig:fig5a}  \includegraphics[width=.9\linewidth,clip]{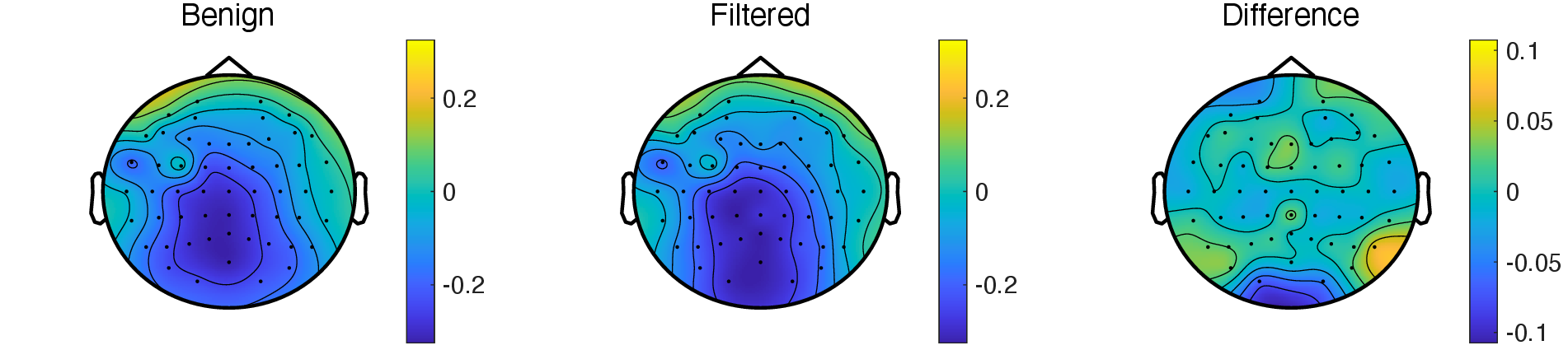}}
\subfigure[]{\label{fig:fig5b}  \includegraphics[width=.9\linewidth,clip]{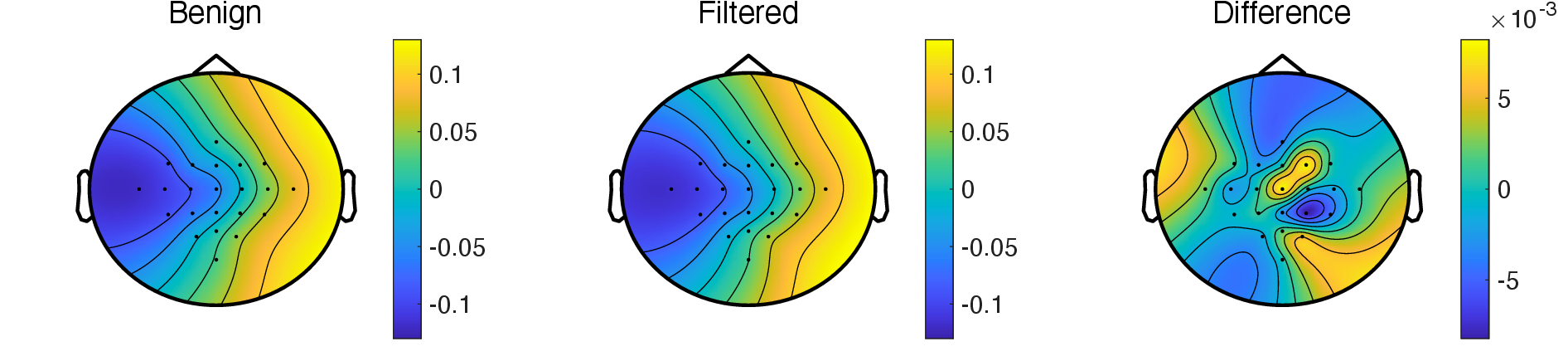}}
\subfigure[]{\label{fig:fig5c}  \includegraphics[width=.9\linewidth,clip]{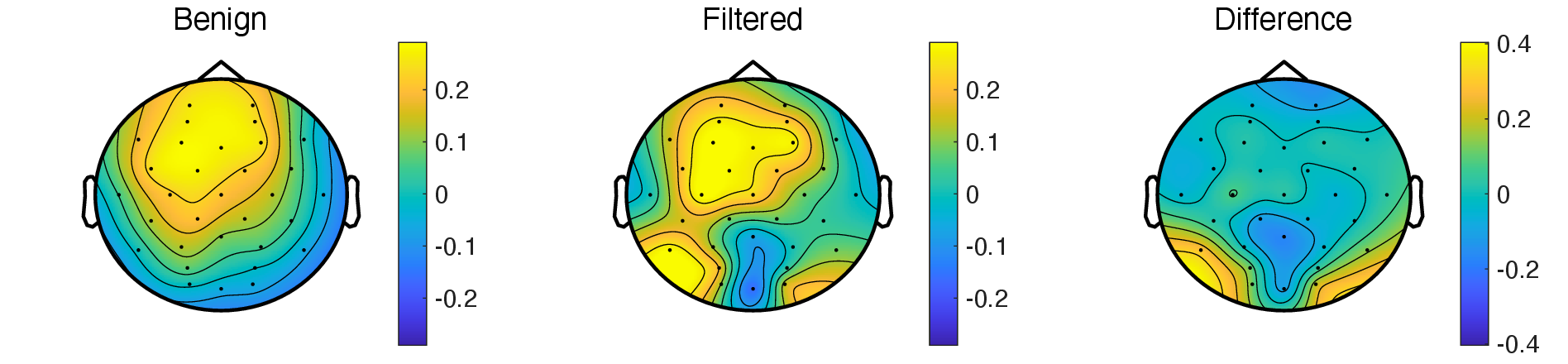}}
\caption{Average topoplots of the benign EEG trials, the EEG trials after adversarial filtering, and their differences. (a) ERN; (b) MI; and, (c) P300.} \label{fig:fig5}
\end{figure*}

Fig.~\ref{fig:fig5} further shows the average topoplots of the benign EEG trials, the EEG trials after adversarial filtering, and their differences. The topoplots from the benign trials and their filtered counterparts were almost identical, making the attack difficult to detect from the topoplots.

\begin{figure}[ht]\centering
\subfigure[]{\label{fig:fig6a}  \includegraphics[width=1.0\linewidth,clip]{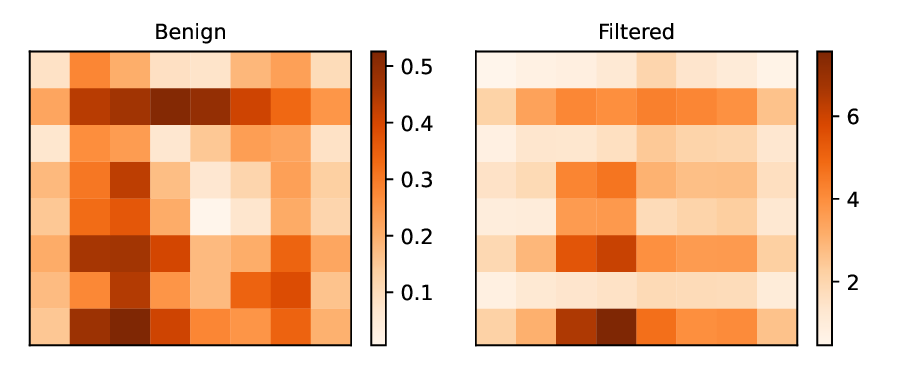}}
\subfigure[]{\label{fig:fig6b}  \includegraphics[width=1.0\linewidth,clip]{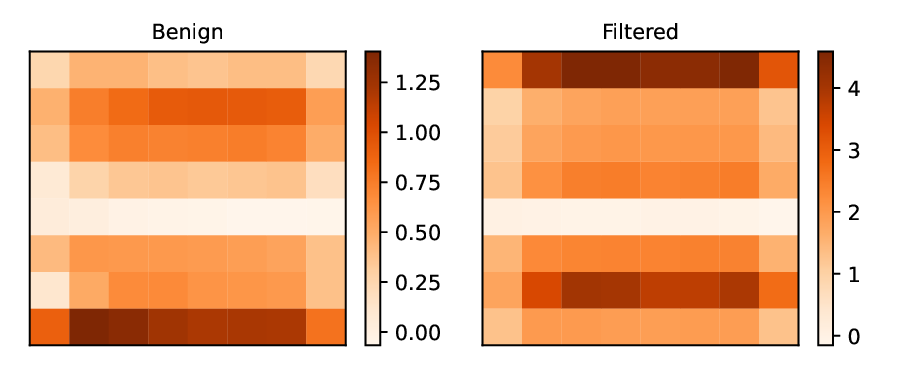}}
\subfigure[]{\label{fig:fig6c}  \includegraphics[width=1.0\linewidth,clip]{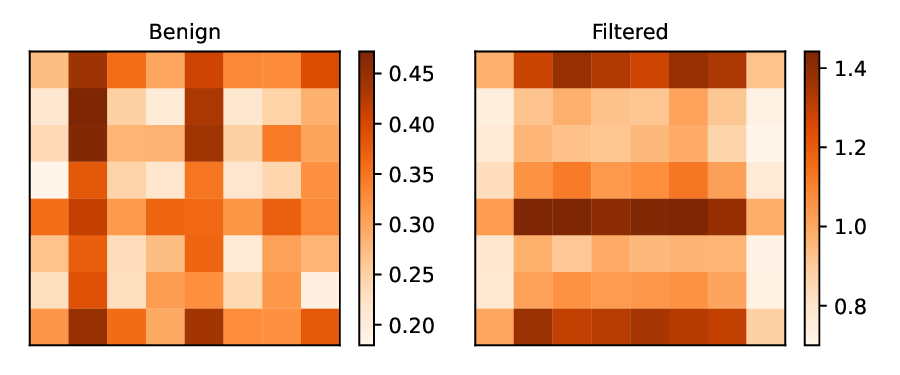}}
\caption{Average EEGNet feature maps of the benign EEG trials and the EEG trials after adversarial filtering. (a) ERN; (b) MI; and, (c) P300.} \label{fig:fig6}
\end{figure}

Fig.~\ref{fig:fig6} visualizes the feature map from the last convolution layer of EEGNet before and after adversarial filtering. Although there was only very small difference between the EEG trials before and after adversarial filtering in the time/frequency domains and on the topoplot, this difference was amplified by the complex nonlinear transformation of EEGNet, and hence the hidden layer feature maps were significantly different. This is intuitive, as otherwise the attack would not succeed.

\begin{table*}[ht]
\centering \setlength{\tabcolsep}{2.8mm}
\caption{BCAs (\%) of two baseline attacks and adversarial filtering based evasion attack.}
\begin{tabular}{c|c|c|c|c|c|c|c}
\toprule
\multirow{3}{*}{Dataset} & \multirow{3}{*}{Victim Model} & \multicolumn{3}{c|}{Within-subject} & \multicolumn{3}{c}{Cross-subject} \\ \cline{3-8}
& & \multicolumn{2}{c|}{Baseline} & Adversarial & \multicolumn{2}{c|}{Baseline} & Adversarial \\ \cline{3-4} \cline{6-7}
& & Clean & Noisy & filter & Clean & Noisy & filter  \\ \midrule
\multirow{4}{*}{ERN} & EEGNet & $73.98\pm1.17$ & $72.99\pm1.16$ & $50.07\pm1.16$ & $65.95\pm0.38$ & $64.78\pm0.61$ & $48.32\pm0.91$ \\
& DeepCNN & $70.15\pm1.52$ & $70.01\pm2.12$ & $48.80\pm1.41$ & $63.61\pm0.43$ & $62.83\pm0.82$ & $49.41\pm0.82$ \\
& ShallowCNN & $69.75\pm1.57$ & $69.38\pm1.67$ & $50.73\pm0.73$ & $64.97\pm0.61$ & $64.73\pm0.75$ & $50.21\pm0.24$ \\
& xDAWN+LR & $70.83\pm2.07$ & $67.72\pm2.05$ & $51.14\pm1.32$ & $62.20\pm0.44$ & $60.69\pm0.73$ & $50.02\pm0.04$ \\ \midrule
\multirow{4}{*}{MI} & EEGNet & $66.45\pm1.28$ & $66.13\pm1.37$ & $25.20\pm0.31$ & $47.94\pm1.14$ & $47.57\pm1.13$ & $25.09\pm0.17$ \\
& DeepCNN & $53.89\pm1.50$ & $53.24\pm1.53$ & $26.00\pm0.84$ & $45.95\pm0.38$ & $45.24\pm0.95$ & $25.03\pm0.09$ \\
& ShallowCNN & $68.57\pm0.93$ & $67.95\pm1.03$ & $25.12\pm0.09$ & $47.54\pm0.36$ & $47.17\pm0.37$ & $25.08\pm0.10$ \\
& CSP+LR & $60.02\pm1.29$ & $52.50\pm0.21$ & $25.00\pm0.00$ & $42.51\pm0.53$ & $41.66\pm0.96$ & $25.01\pm0.03$ \\ \midrule
\multirow{4}{*}{P300} & EEGNet & $80.68\pm0.79$ & $80.57\pm0.55$ & $48.28\pm0.79$ & $63.02\pm0.20$ & $62.98\pm0.26$ & $49.01\pm0.26$ \\
& DeepCNN & $81.50\pm0.75$ & $81.54\pm0.67$ & $47.59\pm0.28$ & $61.75\pm0.43$ & $61.81\pm0.37$ & $48.79\pm0.28$ \\
& ShallowCNN & $80.24\pm0.50$ & $79.92\pm0.58$ & $49.99\pm0.13$ & $63.00\pm0.18$ & $62.94\pm0.26$ & $50.01\pm0.02$ \\
& xDAWN+LR & $75.60\pm0.98$ & $75.31\pm1.05$ & $50.04\pm0.11$ & $59.93\pm0.13$ & $59.97\pm0.25$ & $50.00\pm0.00$ \\ \bottomrule
\end{tabular}
\label{tab:evasion}
\end{table*}

\subsection{Backdoor Attack Performance}

We trained models on the clean training set without any poisoning samples as our baseline. If an EEG trial with the backdoor key can successfully attack the benign machine learning model, then there is no need to poison the training set to create a backdoor in the model.

Table~\ref{tab:backdoor} presents the attack performance of the baseline and the proposed backdoor attack approach. Obverse that:
\begin{enumerate}
	\item The BCAs on baseline and adversarial filtering based backdoor attack were very similar, indicating that adding a small number of poisoning samples to the training set did not significantly affect the machine learning models' classification performance on clean test samples. This ensures that the backdoor attack is not easily detectable.
	\item The baseline ASRs on different models and datasets were very low, indicating that benign machine learning models cannot be easily fooled by test samples with the backdoor key. The ASR differences across different models suggest that different models had different robustness to adversarial filtering.
	\item The ASRs on adversarial filtering based backdoor attacks were significantly higher than the baseline ASRs on all models and datasets, indicating the effectiveness of our proposed backdoor attack approach. The ASRs in within-subject experiments were lower than those in cross-subject experiments, attributed to the difference in training data size between the two experiment scenarios. Even though both experiments poisoned $5\%$ of the training set, within-subject experiments had much fewer training data, reducing the effectiveness of backdoor attacks.
\end{enumerate}

\begin{table*}[htbp]
\centering \setlength{\tabcolsep}{0.7mm}
\caption{BCAs (\%) and ASRs (\%) of baseline attack and adversarial filtering based backdoor attack.}
\begin{tabular}{c|c|c|c|c|c|c|c|c|c}
\toprule
\multirow{3}{*}{Dataset} & \multirow{3}{*}{Victim Model} & \multicolumn{4}{c|}{Within-subject} & \multicolumn{4}{c}{Cross-subject} \\ \cline{3-10}
& & \multicolumn{2}{c|}{Baseline} & \multicolumn{2}{c|}{Adversarial filtering} & \multicolumn{2}{c|}{Baseline} & \multicolumn{2}{c}{Adversarial filtering} \\ \cline{3-10}
& & BCA & ASR & BCA & ASR & BCA & ASR & BCA & ASR \\ \midrule
\multirow{4}{*}{ERN} & EEGNet & $73.98\pm1.17$ & $9.29\pm3.59$ & $73.89\pm1.47$ & $45.91\pm7.53$ & $65.95\pm0.38$ & $20.71\pm2.86$ & $65.08\pm0.79$ & $91.97\pm3.19$ \\
& DeepCNN & $70.15\pm1.52$ & $8.93\pm4.60$ & $68.42\pm1.86$ & $41.38\pm5.09$ & $63.61\pm0.43$ & $19.31\pm2.06$ & $63.53\pm0.66$ & $94.56\pm1.65$ \\
& ShallowCNN & $69.75\pm1.57$ & $6.83\pm2.17$ & $67.89\pm1.76$ & $81.05\pm3.56$ & $64.97\pm0.61$ & $19.15\pm3.15$ & $64.81\pm0.52$ & $96.89\pm1.37$ \\
& xDAWN+LR & $70.83\pm2.07$ & $38.64\pm8.45$ & $69.44\pm2.16$ & $93.57\pm4.44$ & $62.20\pm0.44$ & $12.25\pm4.84$ & $61.64\pm0.47$ & $99.87\pm0.14$ \\ \midrule
\multirow{4}{*}{MI} & EEGNet & $66.45\pm1.28$ & $6.53\pm2.29$ & $66.26\pm2.01$ & $89.76\pm4.38$ & $47.94\pm1.14$ & $10.51\pm7.01$ & $47.31\pm1.03$ & $93.59\pm2.27$ \\
& DeepCNN & $53.89\pm1.50$ & $3.07\pm0.97$ & $52.59\pm1.19$ & $50.98\pm5.33$ & $45.95\pm0.38$ & $11.26\pm4.58$ & $45.66\pm1.07$ & $92.94\pm2.27$ \\
& ShallowCNN & $68.57\pm0.93$ & $4.11\pm2.65$ & $68.88\pm0.89$ & $91.94\pm1.68$ & $47.54\pm0.36$ & $11.48\pm6.56$ & $47.22\pm0.23$ & $97.37\pm0.80$ \\
& CSP+LR & $60.02\pm1.29$ & $4.05\pm3.72$ & $58.50\pm1.00$      & $96.21\pm4.19$ & $42.51\pm0.53$ & $15.45\pm11.77$ & $41.46\pm1.06$ & $98.22\pm1.83$ \\ \midrule
\multirow{4}{*}{P300} & EEGNet & $80.68\pm0.79$ & $4.30\pm0.80$ & $79.26\pm0.60$ & $92.15\pm4.99$ & $63.02\pm0.20$ & $4.59\pm0.67$ & $60.52\pm0.90$ & $98.35\pm1.32$ \\
& DeepCNN & $81.50\pm0.75$ & $3.31\pm0.49$ & $76.85\pm0.75$ & $90.52\pm1.59$ & $61.75\pm0.43$ & $2.99\pm0.68$ & $58.19\pm0.92$ & $89.63\pm3.38$ \\
& ShallowCNN & $80.24\pm0.50$ & $2.94\pm0.45$ & $78.31\pm0.51$ & $96.88\pm0.54$ & $63.00\pm0.18$ & $4.38\pm0.90$ & $61.70\pm0.45$ & $98.04\pm1.03$ \\
& xDAWN+LR & $75.60\pm0.98$ & $7.32\pm2.79$ & $74.15\pm0.69$ & $75.26\pm6.97$ & $59.93\pm0.13$ & $12.28\pm4.82$ & $61.64\pm0.47$ & $99.87\pm0.14$ \\
\bottomrule
\end{tabular}
\label{tab:backdoor}
\end{table*}

\begin{table*}[htbp]
\centering \setlength{\tabcolsep}{4.5mm}
\caption{BCAs (\%) of baseline attacks and evasion attack using adversarial filter generated by different models.}
\begin{tabular}{c|c|c|c|c|c|c} \toprule
\multirow{2}{*}{Dataset} & \multirow{2}{*}{Victim Model} & \multicolumn{2}{c|}{Baseline} & \multicolumn{3}{c}{Generation model} \\ \cline{3-7}
& & Clean & Noisy & EEGNet & DeepCNN & ShallowCNN \\ \midrule
\multirow{3}{*}{ERN} & EEGNet & $65.95\pm0.38$ & $64.78\pm0.61$ & $48.32\pm0.91$ & $50.46\pm1.06$ & $54.10\pm1.23$ \\
& DeepCNN & $63.61\pm0.43$ & $62.83\pm0.82$ & $49.22\pm0.86$ & $49.41\pm0.82$ & $51.60\pm0.69$ \\
& ShallowCNN & $64.97\pm0.61$ & $64.73\pm0.75$ & $54.52\pm0.83$ & $52.33\pm1.26$ & $50.21\pm0.24$ \\ \midrule
\multirow{3}{*}{MI} & EEGNet & $47.94\pm1.14$ & $47.57\pm1.13$ & $25.09\pm0.17$ & $25.23\pm0.24$ & $25.61\pm0.64$ \\
& DeepCNN & $45.95\pm0.38$ & $45.24\pm0.95$ & $25.80\pm0.86$ & $25.03\pm0.09$ & $25.68\pm0.50$ \\
& ShallowCNN & $47.54\pm0.36$ & $47.17\pm0.37$ & $26.28\pm0.98$ & $25.08\pm0.07$ & $25.08\pm0.10$ \\
    \midrule
\multirow{3}{*}{P300} & EEGNet & $63.02\pm0.20$ & $62.98\pm0.26$ & $49.01\pm0.26$ & $49.04\pm0.33$ & $50.69\pm1.73$ \\
& DeepCNN & $61.75\pm0.43$ & $61.81\pm0.37$ & $49.14\pm0.33$ & $48.79\pm0.28$ & $50.74\pm1.65$ \\
& ShallowCNN & $63.00\pm0.18$ & $62.94\pm0.26$ & $50.31\pm0.37$ & $50.33\pm0.42$ & $50.01\pm0.02$ \\ \bottomrule
\end{tabular}
\label{tab:Transferability}
\end{table*}

\subsection{Transferability of the Adversarial Filter}

Table~\ref{tab:Transferability} shows the transferability of the adversarial filter in evasion attacks, i.e., the effectiveness of the adversarial filter generated by one model on other models. We can observe that:
\begin{enumerate}
	\item The adversarial filter generated by one model significantly reduced the BCAs of other models, even though these models had different structures, demonstrating the strong transferability of the adversarial filter.
	\item In general, the lowest BCA was achieved when the machine learning model used to generate the adversarial filter matched the model to be attacked, which is intuitive.
\end{enumerate}

The transferability of the adversarial filter also indicates its applicability in  black-box scenario, where the adversary knows neither the architecture nor the parameters of the machine learning model. In this scenario, the adversary can use any substitute model to obtain an adversarial filter, and then exploits the transferability to attack an unknown machine learning model.

\subsection{Influence of the Trade-off Parameter $\alpha$ in Evasion Attack}

Fig.~\ref{fig:fig6} shows the BCAs of EEGNet and the distortions of test EEG trials under evasion attacks with different trade-off parameters $\alpha\in \{0.1, 1, 10, 50, 100, 200, 500, 1000, 10000\}$). Note that binary search was not used in these experiments. Distortions were calculated as the root mean squared error of the differences between EEG trials before and after adversarial filtering.

\begin{figure*}[htbp]\centering
\includegraphics[width=1.0\linewidth,clip]{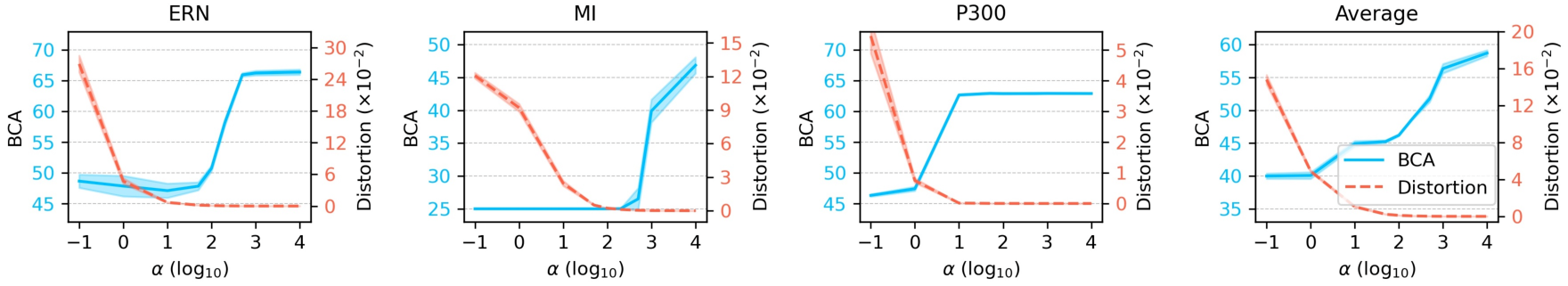}
\caption{BCAs (\%) of EEGNet and the distortions of test EEG trials under evasion attack with different trade-off parameter $\alpha$. The mean and standard deviations were computed from 5 repeats of experiments.} \label{fig:fig7}
\end{figure*}

\begin{figure*}[htbp]\centering
\includegraphics[width=1.0\linewidth,clip]{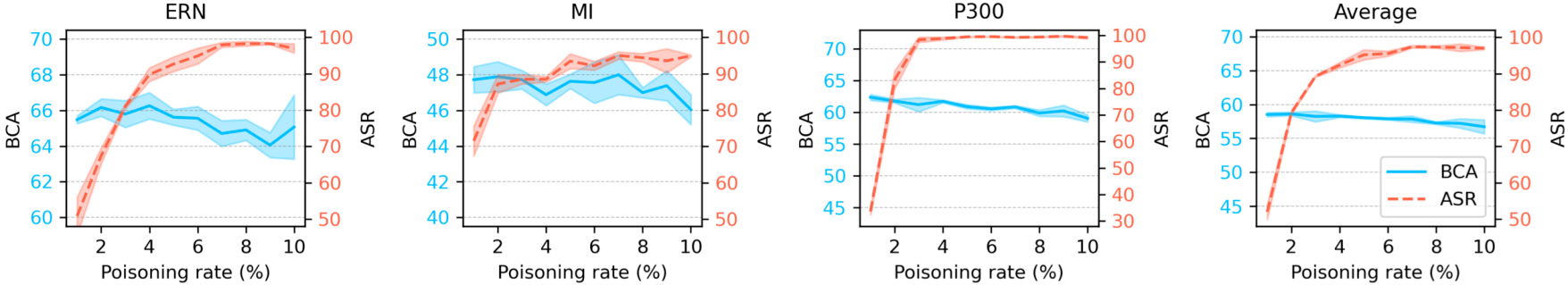}
\caption{BCAs (\%) of poisoned EEGNet and ASRs of the proposed backdoor attack at different poisoning ratios. The mean and standard deviations were computed from 5 repeats of the experiments.} \label{fig:fig8}
\end{figure*}

As $\alpha$ decreased, BCAs decreased and eventually converged to the chance level, whereas distortions increased. This indicated that reducing $\alpha$ enhanced the attack effectiveness but reduced its stealthiness. Therefore, it is necessary to use binary search to find a trade-off parameter that ensures the attack effectiveness while minimizing the distortion.

\subsection{Influence of the Number of Poisoning Samples in Backdoor Attack}

Fig.~\ref{fig:fig7} shows the BCAs of the poisoned EEGNet and ASRs of the proposed backdoor attack when the poisoning ratio (the number of poisoning samples divided by the number of training samples) increased from $1\%$ to $10\%$.

As the poisoning ratio increased, the BCA slightly degraded, whereas the ASR significantly improved. A higher poisoning ratio results in a higher ASR, but also makes the attack easier to detect. On all three datasets, a poisoning ratio of only $1\%$ achieved an average ASR of $50\%$, and a poisoning ratio of $4\%$ was enough to achieve an average ASR of $90\%$.

\section{Conclusions and Future Research} \label{sec:conclusions}

Adversarial attacks to EEG-based BCIs has attracted much research interest recently. This paper has proposed filtering based evasion attack and backdoor attack approaches to EEG-based BCIs, and demonstrated their effectiveness for multiple machine learning models in three different BCI paradigms.

Our future research will:
\begin{enumerate}
	\item Study filtering based adversarial attacks in EEG-based BCI regression problems, such as driver fatigue estimation \cite{drwuFWET2019} and reaction time estimation \cite{drwuRG2017}.
	\item Systematically study the security of BCIs. Existing studies have explored the security of signal processing and machine learning in BCIs separately. However, as a system, studying the overall vulnerability of BCIs may expose more serious security problems.
	\item Develop defense strategies against filtering based adversarial attacks~\cite{drwuFGCS2023}. The ultimate goal of our research is to enhance the security of BCIs, instead of damaging them.
\end{enumerate}

\end{document}